**Title:** Deep Learning Predicts Mammographic Breast Density in Clinical Breast Ultrasound Images

Arianna Bunnell, MS.

**Affiliations:** Information and Computer Sciences, University of Hawai'i at Mānoa, Honolulu, HI, USA. University of Hawai'i Cancer Center, Honolulu, HI, USA.
**Address:** POST Bldg Room 305B; 1680 East-West Rd; Honolulu, HI 96822

Dustin Valdez, MS.

**Affiliations:** University of Hawai'i Cancer Center, Honolulu, HI, USA
**Address:** University of Hawaii Cancer Center; 701 Ilalo St; Suite 522; Honolulu, HI 96813, USA.

Thomas K. Wolfgruber, PhD.

**Affiliations:** University of Hawai'i Cancer Center, Honolulu, HI, USA
**Address:** University of Hawaii Cancer Center; 701 Ilalo St; Suite 522; Honolulu, HI 96813, USA.

Brandon Quon, MPH

**Affiliations:** University of Hawai'i Cancer Center, Honolulu, HI, USA
**Address:** University of Hawaii Cancer Center; 701 Ilalo St; Suite 225; Honolulu, HI 96813, USA.

Kailee Hung, BS.

**Affiliation:** Information and Computer Sciences, University of Hawai'i at Mānoa, Honolulu, HI, USA.
**Address:** POST Bldg Room 306C; 1680 East-West Rd; Honolulu, HI 96822

Brenda Y. Hernandez, PhD

**Affiliations:** University of Hawai'i Cancer Center, Honolulu, HI, USA
**Address:** University of Hawaii Cancer Center; 701 Ilalo St; Suite 239; Honolulu, HI 96813, USA.

Todd B. Seto, MD
**Affiliations:** The Queen's Medical Center
**Address:** The Queen's Health Systems; 1301 Punchbowl Street, Honolulu, HI 96813

Jeffrey Killeen, MD
**Affiliations:** Hawai'i Pacific Health
**Address:** Hawai'i Pacific Health; 55 Merchant St. Honolulu, HI 96813

Marshall Miyoshi
**Affiliations:** Hawai'i Diagnostic Radiology Services
**Address:** Hawai'i Diagnostic Radiology Services (St. Francis); 2230 Liliha Street; Suite 106; Honolulu, HI 96817

Peter Sadowski, PhD.
**Affiliations:** Information and Computer Sciences, University of Hawai'i at Mānoa, Honolulu, HI, USA.
**Address:** POST Bldg Room 306C; 1680 East-West Rd; Honolulu, HI 96822

<u>**Corresponding Author:**</u> John A. Shepherd, PhD.

**Affiliation:** University of Hawai'i Cancer Center, Honolulu, Hawaii, USA.
**Address:** University of Hawaii Cancer Center**;** 701 Ilalo St; Suite 522; Honolulu, HI 96813, USA.

# ABSTRACT

***Background.*** Breast density, as derived from mammographic images and defined by the Breast Imaging Reporting & Data System (BI-RADS), is one of the strongest risk factors for breast cancer. Breast ultrasound (BUS) is an alternative breast cancer screening modality, particularly useful in low-resource, rural contexts. To date, BUS has not been used to inform risk models that need breast density. In this study, we explore an artificial intelligence (AI) model to predict BI-RADS breast density category from clinical BUS imaging.

***Methods.*** We compared deep learning methods for predicting breast density directly from BUS imaging, as well as machine learning models from BUS image statistics alone. The use of AI-derived BUS breast density as a breast cancer risk factor was compared to clinical BI-RADS breast density. The BUS data were split by individual into 70/20/10% groups for training, validation, and held-out testing for reporting results.

***Findings.*** 405,120 clinical BUS images from 14,066 women were selected for inclusion: 9,846 training (302,574 images), 2,813 validation (11,223), and 1,406 testing (4,042). The strongest AI model achieves AUROC 0·854 and outperforms all image statistic-based methods. In risk prediction, age-adjusted AI BUS and clinical breast density predict 5-year breast cancer risk with 0·633 and 0·637 AUROC, respectively.

***Interpretation.*** BI-RADS breast density can be estimated from BUS imaging with high accuracy. The AI model provided superior estimates to other machine learning approaches. Furthermore, we demonstrate that age-adjusted, AI-derived BUS breast density is predictive of 5-year breast cancer risk in our population.

***Funding.*** National Cancer Institute.

**INTRODUCTION**

Other than age, mammographic breast density is one of the strongest risk factors for breast cancer. Breast density has been extensively studied over the past 30 years and is included in many breast cancer risk models.[1-3] A recent meta-analysis showed that having extremely dense breasts increased lifetime breast cancer risk by 2·11 times over women with scattered dense breast tissue (BI-RADS density B), even when adjusted for BMI and age.[4] High breast density lowers the sensitivity of mammography leading to a federal mandate to report breast density to all women undergoing screening mammography.[5] Thus, many women have direct knowledge of the breast density and can apply it to readily available risk models. Women at high risk of developing breast cancer have multiple options for lowering their risk including lifestyle choices,[6] and chemoprevention.[7-9] However, there are many parts of the world that are either too resource-limited or too remote to have access to screening mammography programs. Without mammography, clinical breast density has not been available for comprehensive breast cancer risk assessment.

Breast ultrasound (BUS) is a sensitive method for detecting breast cancer and is broadly used around the world as primary screening in lower resource settings,[10-12] secondary screening for women with dense breast tissue,[13-15] and follow-up adjuvant imaging.[16,17] However, mammography is preferred, when available, due to is lower false-positive rate.[18] Ideally, breast density would be available from accessible breast imaging technologies, like BUS, with a calibration equivalent to mammographic density.

Image contrast in B-mode BUS is based on the boundaries where density changes. See *Figure 1* for examples of matched BUS and mammography images at different mammographic densities. These boundary features have been investigated in the past to derive density-like measures. Jud et al.[19] used linear regression of variable defined by the grey-level values normalized into 16 bins. Their model was able to estimate percent mammographic density with an accuracy of $R^2 = 0·67$ but suffered from calibration issues in external datasets.[20]

In the past decade, there has been a revolution in machine learning and convolutional neural deep learning networks with substantial literature on predicting mammographic breast density from mammograms using artificial intelligence (AI)[21-23] and deriving breast cancer risk from mammograms directly.[24-27] The purpose of this study is to explore the use of AI to estimate breast density from BUS images. We further explore the use of our AI-derived BUS breast density in place of mammographic density in a simple risk model.

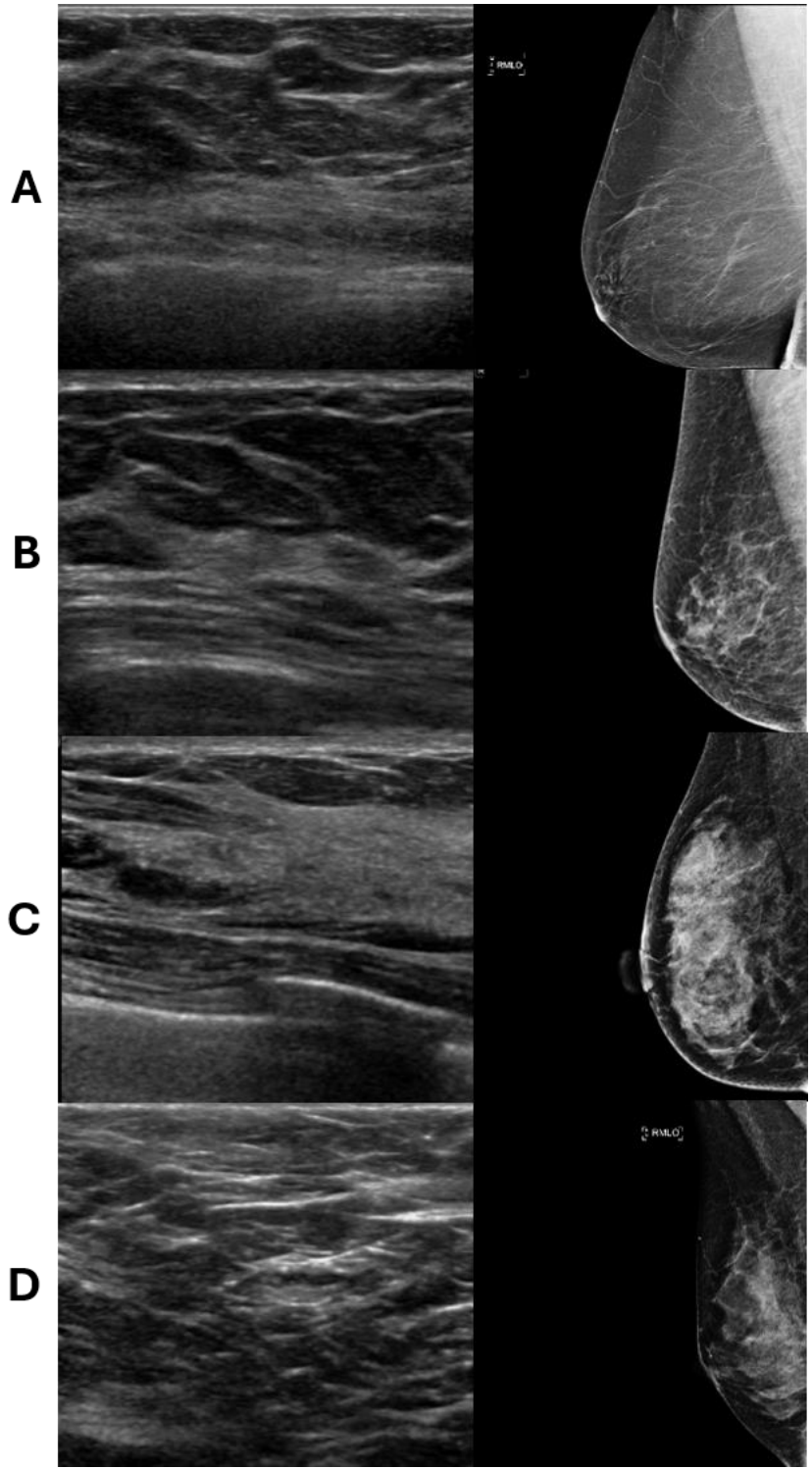

*Figure 1.* Example clinical breast ultrasound (left) and right mediolateral oblique digital 2D mammography (right) images, selected from the dataset included in this work. Image sets are labeled with their BI-RADS breast density category. Images have been cropped for ease of display.

**METHODS**

Using retrospective, clinical BUS images, we constructed a deep learning model to predict BI-RADS mammographic density. We compared our deep learning model to a previously-derived machine learning method using grey-level image histograms. We then compare the predictive ability of clinical, AI-derived and image grey-level histogram-based BUS estimates of breast density in estimating 5-year breast cancer risk.

**Study sample**

All women included in this study participated in either screening or diagnostic BUS imaging in the Hawaiʻi and Pacific Islands Mammography Registry (HIPIMR; see Supplement). Women were selected for inclusion if they met all the following criteria: (a) had at least one negative screening 2D mammography visit; (b) had a negative, benign, or probably benign BUS visit within one year of their mammograms; (c) had a clinical BI-RADS breast density; (d) had the standard four views captured at their screening mammogram; and (e) had no history of breast cancer prior to screening mammogram (See Supplement for complete inclusion/exclusion criteria). From these, cases were defined as women diagnosed with invasive breast cancer at least 6 months and no more than 5 years from their extracted BUS exam. Controls were selected from women who did not develop cancer. See *Figure 2* for the complete data flow diagram.

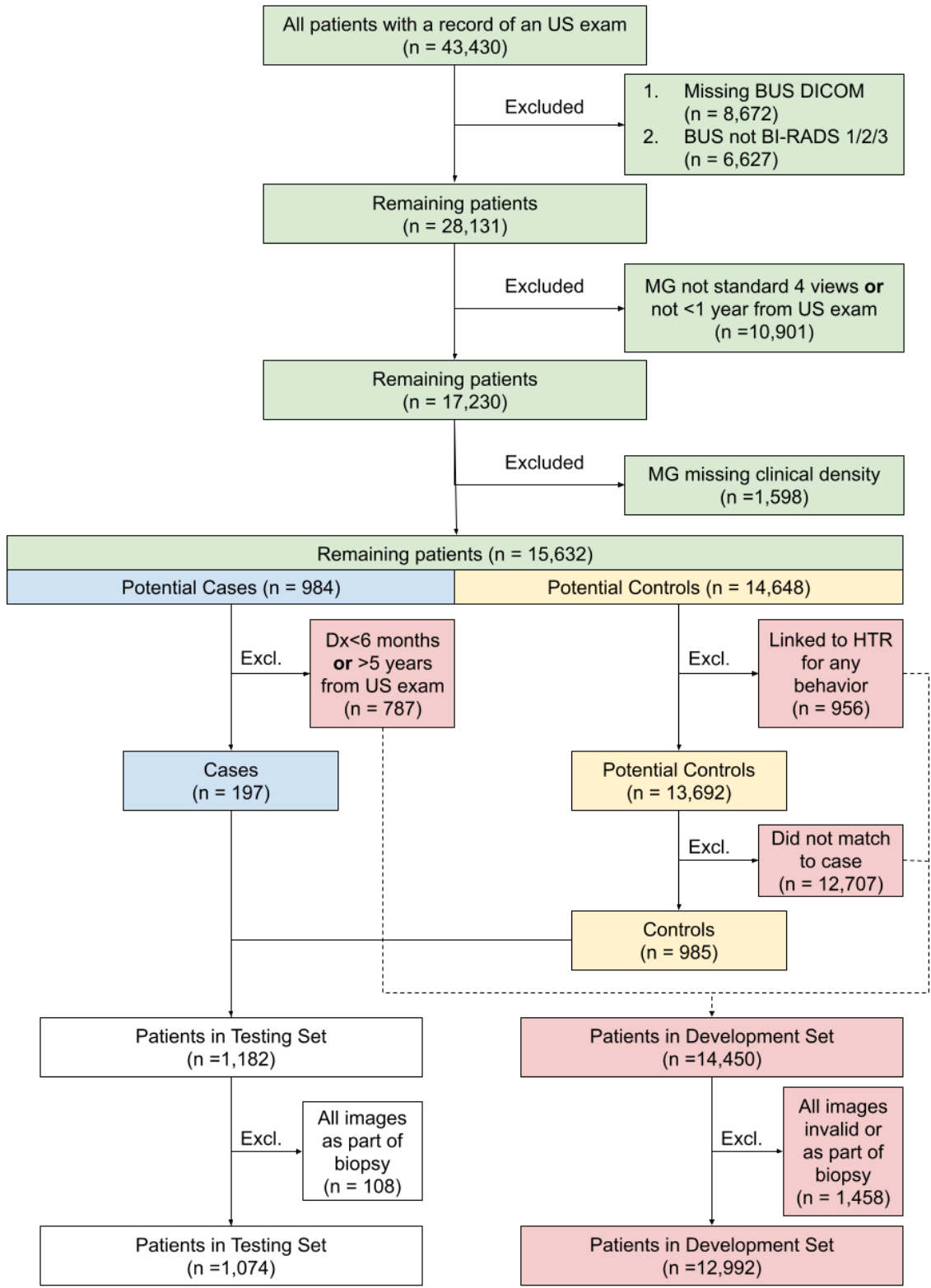

*Figure 2*. Flow diagram showing data selection process from the HIPIMR. Cases and controls were selected from the same pool of 15,632 patients with matched mammograms with density records and BUS exams. 197 cases and 965 controls, matched on birth year, were selected for the testing set. After exclusion of invalid images, 1,074 women remained. The remaining 12,992 eligible women with valid images were allocated to the development set. Dx = diagnosis; MG = mammogram; BUS = breast ultrasound; Excl. = excluded; HTR = Hawaiʻi Tumor Registry.

**Imaging data preparation**

Clinical BUS data are highly noisy data, with many artifacts such as Color Doppler blood flow highlighting, sonographer text annotations, and lesion caliper markers which can interfere with AI model learning. We implement a cleaning pipeline developed in-house called BUSClean[28] to remove these artifacts and standardize BUS images converted from DICOM to PNG.

**Data splitting**

The development dataset was split by woman, stratified by clinical BI-RADS breast density category, with 80% assigned to the training data set and 20% assigned to the validation split for hyperparameter tuning and early stopping determination. The case-control matched testing set was reserved for performance evaluation. Each woman and her images are only present in a single data split. We duplicated the validation dataset to create *curated* and *uncurated* validation sets. In brief, the *curated* validation set has extensive preprocessing applied to remove clinical artifacts, while the *uncurated* validation set had only invalid scans removed and scans with multiple views split. See Supplement for details.

**Deep learning model**

Three deep learning models were trained in PyTorch[29] using three distinct ImageNet-pretrained architectures: DenseNet121[30], ViT-B/32[31], and ResNet50[32], selected to represent a diverse range of deep learning architecture designs. Hyperparameters for each were optimized over 25 trials using the TPESampler in Optuna[33]. *Supplemental Table 1* displays the search space. The best-performing model and hyperparameters on the curated validation set was selected as the final architecture and retrained (see *Supplement* for training configuration). The final deep learning architecture was selected based on performance on the curated validation set.

following their screening examination, matched to cases 5:1 on year of screening mammography

**Image histogram models**

Prior work in estimating percentage mammographic breast density from B-mode BUS imaging from Jud et al. makes use of gray-level image histograms with equally-sized intervals in a linear regression model.[19] We implement a version of their method adapted for the categorical BI-RADS breast density measure (see Supplement). Logistic regression, random forest, and multi-layer perceptron (MLP) models were constructed using the gray-level image histogram intervals as features to predict BI-RADS categorical breast density.

**Statistical analysis**

Statistical analysis was undertaken in two steps. The first step demonstrates AI model accuracy in predicting breast density category from BUS. The second step demonstrates utility of predicted BUS BI-RADS breast density category in breast cancer risk modelling. See Supplement for the implementation details of both steps. To assess AI model performance in predicting breast density from BUS, all models are evaluated by their micro-averaged AUROC over all density categories with 95% confidence intervals estimated using DeLong's method[34,35] on the testing set. We use micro-averaged AUROC for increased stability under class imbalance. We present AUROC on the overall testing set as well by BUS machine manufacturer, binned patient age at BUS exam, cancer status, and BUS exam BI-RADS category. We additionally compute Kendall's τ-b[36] between clinical density and predicted density from BUS to compare agreement. As a comparator, we computed mammographic density using Geras et al.[21] on all included women and computed Kendall's τ-b between these predictions and clinical density using pre-existing interpretation guidelines.[37] To demonstrate utility in cancer risk prediction, we constructed logistic regression models of breast cancer risk from age at BUS exam and mammographic breast density. Effect of breast density category and standardized age was compared via computation of odds ratios with 95% Wald confidence intervals.

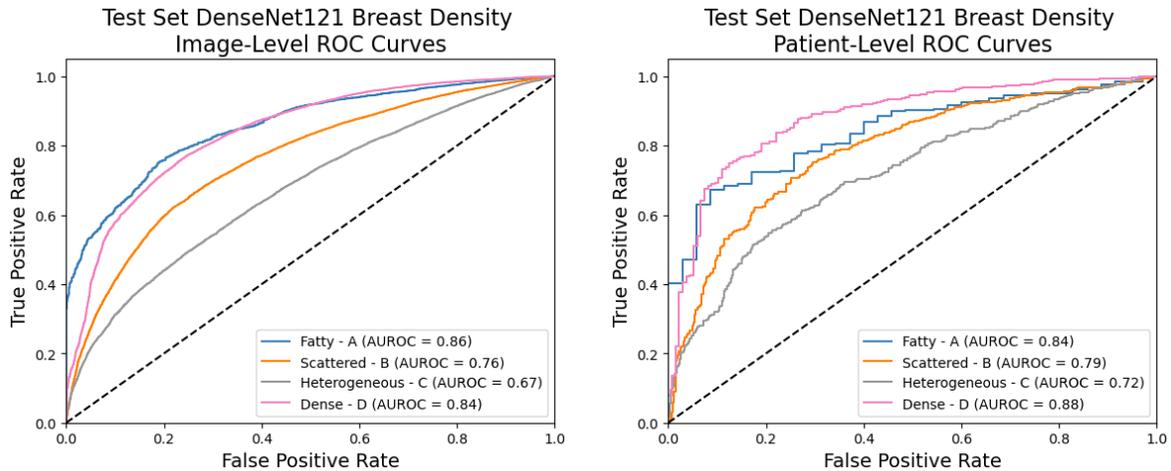

*Figure 3*. BUS AI model receiver operating characteristic (ROC) curves for the image-level predictions (left) and when aggregated through arithmetic averaging to patient-level (right).

**RESULTS**

**Study sample characteristics**

The characteristics of all 14,066 women (15,632 identified for inclusion from the HIPIMR) included in the study (mean age at BUS exam ± standard deviation, 53 ± 12 years) are described in *Table 1*. See the Supplement for complete inclusion/exclusion criteria-specific counts. 405,120 total BUS images were included in this study

**Deep learning model**

Based on superior performance in the curated validation set, DenseNet121 was chosen as the final deep learning model architecture (see *Supplemental Table 1*). Performance with confidence interval estimates on the complete testing set, as well as subgroups of interest, are reported in *Table 2*. DenseNet121 identified patients with BI-RADS breast density category A, B, C, and D from all

*Table 1.* Study sample characteristics. See the supplement for counts of the *uncurated* validation set. BUS = breast ultrasound; MG = mammogram; Dx = diagnosis; IQR = inter-quartile range; SD = standard deviation.

|  | Overall | Training Set | Curated Validation Set | Testing Set |
|---|---|---|---|---|
| **Women, N** | 14,066 | 10,393 | 2,593 | 1,074 |
| Women with benign findings, N (%) | - | - | - | 896 (83.4) |
| Women with malignant findings, N (%) | - | - | - | 178 (16.6) |
| Mean age at US, years (SD) | 53.4 (11.9) | 53.4 (11.9) | 53.3 (12.0) | 53.8 (12.1) |
| Mean age at MG, years (SD) | 53.4 (11.9) | 53.4 (11.9) | 53.2 (12.0) | 53.8 (12.1) |
| Mean age at diagnosis, years (SD) | - | - | - | 60.4 (11.4) |
| Women with fatty/A breasts, N (%) | 478 (3.4) | 344 (3.3) | 99 (3.8) | 35 (3.3) |
| Women with scattered/B breasts, N (%) | 5,486 (39.0) | 4,088 (39.3) | 987 (38.1) | 409 (38.1) |
| Women with heterogeneous/C breasts, N (%) | 6,291 (44.7) | 4,615 (44.4) | 1,178 (45.4) | 494 (45.9) |
| Women with dense/D breasts, N (%) | 1,811 (12.9) | 1,346 (13.0) | 329 (12.7) | 136 (12.7) |
| Menopausal women, N (%) | 4,365 (31.0) | 3,210 (30.9) | 806 (31.1) | 345 (32.1) |
| Premenopausal women, N (%) | 480 (3.4) | 347 (3.3) | 83 (3.2) | 49 (4.6) |
| Unknown menopausal status, N (%) | 9,221 (65.6) | 6,836 (65.8) | 1,704 (65.7) | 680 (63.3) |
| Median gap btw. MAM & US, days (IQR) | 5.0 (20.0) | 5.0 (20.0) | 4.0 (19.0) | 6.0 (19.8) |
| Median gap btw. MAM & Dx, days (IQR) | - | - | - | 822.0 (816.8) |
| Median gap btw. US & Dx, days (IQR) | - | - | - | 808.5 (808.5) |
| **Images, N** | 405,120 | 302,574 | 69,842 | 28,616 |
| Mean no. of images per woman, N (SD) | 28.8 (22.9) | 29.1 (23.1) | 26.9 (21.8) | 26.6 (21.9) |
| Images with benign findings, N (%) | - | - | - | 23,901 (83.5) |
| Images with malignant findings, N (%) | - | - | - | 4,715 (16.5) |
| US BI-RADS 1 images, N (%) | 58,428 (14.4) | 43,163 (14.3) | 10,856 (15.5) | 4,042 (14.1) |
| US BI-RADS 2 images, N (%) | 238,268 (58.8) | 177,280 (58.6) | 41,313 (59.2) | 17,419 (60.9) |
| US BI-RADS 3 images, N (%) | 108,424 (26.8) | 82,131 (27.1) | 17,673 (25.3) | 7,155 (25.0) |
| Images of fatty/A breasts, N (%) | 11,525 (2.9) | 8,416 (2.9) | 2,223 (3.2) | 886 (3.1) |
| Images of scattered/B breasts, N (%) | 151,015 (37.6) | 115,394 (38.1) | 25,794 (36.9) | 9,827 (34.3) |
| Images of heterogeneous/C breasts, N (%) | 180,927 (45.1) | 135,622 (44.8) | 31,941 (45.7) | 13,364 (46.7) |
| Images of dense/D breasts, N (%) | 57,565 (14.4) | 43,142 (14.2) | 9,884 (14.2) | 4,539 (15.9) |
| Images on PHILIPS System, N (%) | 363,308 (89.7) | 271,905 (89.9) | 64,204 (91.9) | 24,687 (86.3) |
| Images on ATL System, N (%) | 25,297 (6.2) | 17,877 (5.9) | 3,723 (5.3) | 3,208 (11.2) |
| Images on SIEMENS System, N (%) | 12,617 (3.1) | 9,795 (3.2) | 1,495 (2.1) | 443 (1.5) |
| Images on TOSHIBA System, N (%) | 3,763 (0.9) | 2,862 (0.9) | 420 (0.6) | 278 (1.0) |
| Images on GE System, N (%) | 126 (0.0) | 126 (0.0) | 0 (0.0) | 0 (0.0) |
| Images on ALOKA System, N (%) | 9 (0.0) | 9 (0.0) | 0 (0.0) | 0 (0.0) |

other categories with AUROC 0·84, 0·79, 0·72, and 0·88 respectively on the unseen test set. *Figure 3* displays receiver operating characteristic curves for the DenseNet121 on the unseen test set, separated by ground truth mammographic density category. In comparison of performance on subgroups (see *Supplemental Table 4*), non-overlapping confidence intervals were observed between patients imaged Toshiba machines and other BUS machine manufacturers, with higher performance in Toshiba machines and patients under 50 years old and over 65 years old, with

*Table 1.* Overall performance statistics for the DenseNet121, logistic regression, and random forest models. See *Supplemental Table 4 for* subgroup performance.

| | | AUC* (95% C.I.) | by BI-RADS breast density category | | | |
|---|---|---|---|---|---|---|
| | | Overall | Fatty/A | Scattered/B | Heterogen./C | Dense/D |
| DenseNet121 | Per-image | 0.840 (0.838, 0.842) | 0.857 (0.847, 0.867) | 0.761 (0.755, 0.766) | 0.673 (0.667, 0.679) | 0.839 (0.833, 0.846) |
| | Per-patient | 0.854 (0.842, 0.866) | 0.841 (0.790, 0.891) | 0.787 (0.759, 0.815) | 0.719 (0.689, 0.749) | 0.879 (0.848, 0.910) |
| Logistic Reg | Per-image | 0.771 (0.768, 0.774) | 0.594 (0.574, 0.613) | 0.607 (0.601, 0.614) | 0.584 (0.577, 0.590) | 0.583 (0.574, 0.592) |
| | Per-patient | 0.795 (0.781, 0.809) | 0.642 (0.552, 0.732) | 0.654 (0.620, 0.687) | 0.621 (0.588, 0.654) | 0.668 (0.619, 0.716) |
| Random Forest | Per-image | 0.759 (0.756, 0.762) | 0.604 (0.586, 0.623) | 0.597 (0.590, 0.604) | 0.540 (0.533, 0.546) | 0.625 (0.617, 0.634) |
| | Per-patient | 0.794 (0.780, 0.808) | 0.698 (0.620, 0.776) | 0.639 (0.605, 0.673) | 0.575 (0.541, 0.609) | 0.727 (0.680, 0.774) |

*AUC is a micro-averaged one vs. rest AUROC over all density categories, except for density subgroups. For density subgroups, one vs. rest AUC for each category is shown. 95% confidence intervals are calculated using DeLong's method.[1,2]

higher performance in younger patients. Performance between all other subgroups was found to be non-significantly different. *Supplemental Figure 1* shows example saliency maps highlighting the areas of the BUS scans which may be of particular importance for the deep learning model's predictions on the unseen test set, computed using integrated gradients.[38,39]

**Image histogram models**

Performance results for the baseline random forest and logistic regression models (see *Supplemental Table 2* for model selection and tuning details) are presented in Error! Reference source not found.. Logistic regression from the gray-level image histograms identified patients in the held-out testing set with BI-RADS breast density A, B, C, and D from all other categories with AUROC 0·642, 0·654, 0·621, and 0·668, respectively. The image histogram random forest model identified patients in the held-out testing set with BI-RADS breast density A, B, C, and D from all other categories with AUROC 0·698, 0·639, 0·575, and 0·727, respectively. 95% confidence intervals for performance overlap between linear (logistic regression) and nonlinear (random forest) modelling from the gray-level image histograms in every BI-RADS breast density category.

**Agreement to Clinical Density**

Kendall's τ-b between predicted BUS density and clinical density was found to be 0·47 (moderate agreement). Predicted BUS density was calculated per-patient by taking an arithmetic mean over the predicted class with the largest probability from each image and rounding to the nearest integer. Kendall's τ-b between mammographic density from Geras et al.[21] and clinical density was found to be 0·60 (moderate agreement). Predicted mammographic density from Geras et al.[21] was defined as the predicted class with the largest probability.

**Cancer Risk Modeling**

The benchmark for age-adjusted AI-derived BI-RADS breast density performance was performance of age-adjusted clinical BI-RADS breast density scores, assigned by the examining radiologist at the time of each woman's matched mammogram. We additionally computed the predictive power of age alone. The age-adjusted, clinical, BI-RADS breast density category predicted cancer occurrence with 0·637 AUROC (95% confidence interval (0·594, 0·679)). The BUS AI density model BI-RADS category predicted cancer occurrence with AUROC 0.633 (0·590, 0·676). Standardized age alone predicted cancer occurrence with AUROC 0·628 (0·585, 0·671). Odds ratios (ORs) for the individual explanatory variables are presented in **Supplemental Table 3.** The ORs with 95% confidence intervals for the extremely dense/D (class B/scattered as reference) clinical and BUS AI density were 1·54 (1·31, 1·81) and 1·50 (1·28, 1·76), respectively.

Additional experiments modeling cancer risk from clinical mammographic breast density in an age-matched cohort resulted in an AUROC of 0·546 (0·478, 0·614), supporting the observed result in the non-age-matched cohort that clinical breast density did not provide additional predictive power. Cases in the age-matched cohort were the same cases included in the previously-defined testing set. Controls were re-matched and selected from the previously-defined validation and testing set. BUS density was not computed for this group.

**DISCUSSION**

We found that BI-RADS breast density can be estimated through machine learning methods from BUS images and that our deep learning model outperformed other approaches. We also found that the top performing BUS breast density estimate had comparable performance or at least was not inferior to clinical BI-RADS breast density for predicting 5-year breast cancer risk in an unseen test set. Furthermore, agreement to clinical density is shown to be similar to an open-source method for predicting density from the mammogram directly.[21] We found that the deep learning model had similar performance in subgroups of BUS machine make and model, patient age, cancer status, and diagnostic BI-RADS status. Given this robustness, breast density derived from our deep learning model should be broadly applicable to clinical situations globally, particularly in remote- and low-resource areas.

Limited prior work has investigated the prediction of mammographic breast density from ultrasound. Tissue speed-of-sound (SoS) using a dedicated non-diagnostic probe has been proposed for measuring mammographic breast density.[40] It was found that non-dense (BI-RADS classes A & B) could be predicted from dense (BI-RADS C & D) breasts with AUROC = 0·887. However, the application of a dedicated BUS device for breast density has limited clinical applications or benefit over being able to get BUS from a clinical hand-held system. Our method can identify patients with extremely dense breasts (class D) with similar performance of AUROC 0·879. When dichotomized, our method identified dense (BI-RADS C & D) from non-dense (BI-RADS A & B) women with AUROC 0·823 (0·797, 0·848). Mammographic percent density has also been measured using SoS on a SoftVue (Delphinus Medical Technologies; MI, USA) ultrasound tomosynthesis system with an accuracy of $R^2 = 0·96$.[41-43] Ultrasound tomosynthesis systems are highly specialized equipment with high installation and maintenance costs which may not be appropriate for low-resource and rural areas. Handheld BUS is most directly applicable to rural and resource-limited scenarios due to portability and low cost. Additional non-mammographic methods of measuring breast density have also been explored, including microwave breast imaging,[44] dual X-ray absorptiometry,[45,46] optical spectroscopy,[46] and bioimpedance.[47,48] These methods, no matter their performance, are less desirable than integrating breast density into clinical BUS since they do not provide cancer detection in addition to density assessment the way clinical BUS does. If used in breast cancer screening, these alternative methods would need to be coupled with BUS or mammography to provide both detection and risk assessment.

Alternative BUS imaging-based biomarkers of breast cancer risk have been explored other than breast density. Breast parenchymal pattern from ultrasound measures the distribution of fat and ductal tissue in the breast. It can be classified by a breast radiologist into four categories: ductal, heterogeneous, mixed, and fibrous and has been found to be associated with breast cancer risk in the Chinese population.[49,50] Another biomarker of breast cancer risk identified from breast ultrasound imaging is glandular tissue component. Glandular tissue component is also a four-category classification which measures the degree of lobular involution observed in *screening*

BUS.[51] High glandular tissue component is associated with increased risk of breast cancer in women with extremely dense (BI-RADS D) breasts.[51,52] It may be that these measures have complementary risk information to breast density but this has yet to be explored. We pursued breast density since it may be integrated into existing breast cancer risk models such as the BCSC, Tyrer-Cuzick and others. Background echotexture, as defined by the ACR, is a three-category classification which has been found to be associated with both mammographic density and parity, known risk factors for breast cancer.[53] AI combined with automated breast ultrasound (ABUS) performed with AUROC 0·979 for identifying background echotexture categories,[54] however ABUS may be prohibitively costly for remote and low-resource rural areas.

The performance of the derived breast risk model using age alone was not significantly different from models with *either* clinical or BUS-derived density, limiting the insights we can draw about the added value of our measure in clinical breast cancer risk models. The typical effect has been found to be 2·1 to 2·3 times the relative risk for women with extremely dense (BI-RADS D) breasts over women with scattered density (BI-RADS B).[55,56] The lack of observed effect of BI-RADS breast density in our study may be attributed to the imaging being from diagnostic visits. BUS is not a primary screening modality in Hawaiʻi, and all exams included in this study are diagnostic exams requested after an initial mammogram. The characteristics of the diagnostic BUS population are likely different from those of a screening mammography population, on which most breast cancer risk evaluation is based. Future validation is needed in a population screened with BUS to verify our findings and confirm clinical utility of BUS-derived breast density in traditional breast cancer risk models.

Although our findings are promising, there are limitations to consider. First, there were relatively few women present in the lowest density category A (4%). This imbalance may be due to the ethnic breakdown of the population of the HIPIMR, with an estimated 60% of women in Hawai'i identifying as Asian (alone or in combination with another race) in 2023.[57] Asian women have been found to have denser breasts than women of other races.[58,59] Future work will explore sample enrichment with category A women. Another limitation of this work is that our risk model only consisted of age and breast density. Inclusion of other breast cancer risk factors may provide a more comprehensive assessment and validation of our method.

We conclude that estimation of categorical, BI-RADS breast density is possible from BUS imaging with acceptable accuracy for use in breast cancer risk assessment models. In cancer risk models, the performance of our measure of breast density indicates an association to cancer risk on par with clinical density in our dataset.

**EVIDENCE BEFORE THIS STUDY**

Age-adjusted mammographic breast density is one of the strongest risk factors for breast cancer and several artificial intelligence (AI) approaches exist for automatic breast density estimation from digital mammography imaging. When using breast ultrasound for primary screening, such as in low-resource medical contexts, breast density is not available. Lack of mammographic breast density limits the breast cancer risk evaluation which can be done. Prior work has shown that breast density can be estimated from modalities other than mammography, with varying levels of success. In ultrasound, speed-of-sound using specialized hardware or methods using automated/3D ultrasound have been the most successful. These methods have limited applicability in low-resource contexts (for ABUS) or require specialized hardware (speed of sound).

**ADDED VALUE OF THIS STUDY**

Our study proposes an AI method for estimation of BI-RADS mammographic breast density category from handheld breast ultrasound imaging and demonstrates the efficacy of breast density estimation from ultrasound in cancer risk assessment. This method could be used to provide more complete breast cancer risk evaluation in rural and low-resource areas where mammography is not available, or in any context where women are primarily being screened with breast ultrasound.

**IMPLICATIONS OF ALL THE AVAILABLE EVIDENCE**

Breast ultrasound imaging contains information about mammographic breast density. Our study shows that BI-RADS mammographic breast density can be accurately estimated from clinical breast ultrasound data using artificial intelligence. Estimated breast density may be useful in performing breast cancer risk assessment in rural and low-resource areas, where mammography may not be available.

**LIST OF ABBREVIATIONS**

BUS = breast ultrasound

AUROC = area under the receiver operating characteristic curve

AI = artificial intelligence

BI-RADS = breast imaging-reporting and data system

ABUS = automated breast ultrasound

HIPIMR = Hawai‘i and Pacific Islands Mammography Registry

**DECLARATIONS**

**Ethics approval and consent to participate.** This study was approved by the WCG IRB (Study Number 1264170).

**Consent for publication.** Not applicable.

**Data sharing statement.** The data that support the findings of this study are available from the Hawaiʻi and Pacific Islands Mammography Registry (HIPIMR), but restrictions apply to the availability of these data. Deidentified data can be requested for research use at any time through the HIPIMR's website at https://hipimr.shepherdresearchlab.org/. All code for running the trained models, creating the gray-level comparator method, and evaluation is available at https://github.com/hawaii-ai/bus-density.

**Competing interests.** The authors declare no conflicts of interest or financial interest.

**Funding.** This study was funded by NCI Grant 5R01CA263491-02.

# Supplement

**HIPIMR description:** The HIPIMR collects data from three clinical partners: Clinic 1 is a nonprofit healthcare network (comprising four medical centers); Clinic 2 is a private nonprofit tertiary hospital; and Clinic 3 is a diagnostic medical imaging center from 2009-2023 on the island of Oʻahu. The HIPIMR is matched to the Hawaiʻi Tumor Registry for retrieval of biopsy- or surgery-confirmed cancer labels. All included mammograms were taken on Hologic Selenia (65.0%) or Selenia Dimensions (35.0%) machines. 89.7% of included BUS images were captured on a Philips Medical Systems IU22 system.

**Inclusion/exclusion criteria:** Women were selected for inclusion if they: (a) had at least one negative screening 2D mammography visit; (b) had a negative, benign, or probably benign (BI-RADS 1, 2, or 3) BUS visit within one year of the recorded screening mammography visit; (c) had a clinical BI-RADS breast density recorded at their screening mammogram; (d) had the standard four views (left craniocaudal, right craniocaudal, left mediolateral oblique, and right mediolateral oblique) captured at their screening mammogram; and (e) had no history of breast cancer prior to screening mammogram. Women with missing mammographic density categories, BI-RADS scores, or imaging were excluded. Male patients were also excluded. Missing covariate data were not imputed but reported as missing. Case-control matching was performed on the testing set only.

**Curated and uncurated validation sets:** The curated validation and testing sets had the complete preprocessing pipeline applied, while the training and uncurated validation sets had only invalid scans removed and dual-view scans split. Training performance was monitored on both the curated and uncurated validation sets as an indicator for the presence of model dependence on spurious correlations with scan artifacts.

**Deep learning architecture modifications:** To conform with model architectures developed on three-channel RGB images, scans are cropped to 224 by 224 pixels (images measuring less than 224 pixels on either dimension have their smaller edge resized to 224 prior to cropping), per-image min-max normalized, and fed through a single convolutional layer with a 1-by-1 kernel and three filters before being passed to an ImageNet-pretrained architecture (4). The final fully-connected layer of the architecture is augmented with random dropout.

*Table 1.* Hyperparameter search space for all Optuna trials for all deep learning architectures tested. DenseNet121 was selected based on clean validation performance. Architecture weights were 25 trials using the TPESampler in Optuna (2).

| Architecture | Chosen Value | | | | | Dirty Validation AUC[‡] | Clean Validation AUC[‡] |
| --- | --- | --- | --- | --- | --- | --- | --- |
| | Learning Rate [1e-7, 1e-3] * | Batch Size {64, 128, 256, 512} | Apply ColorJitter in Training {True, False} | First Trainable Backbone Unit {0, 1, 2, 3, 4} [†] | Penultimate Dropout Proportion [0.0, 0.1, …, 0.9] | | |
| ResNet50 | 1.19e-7 | 512 | False | 0 | 0.9 | 0.861 | 0.805 |
| DenseNet121 | 5.28e-7 | 256 | False | 4 | 0.9 | 0.864 | 0.840 |
| ViTB32 | 5.46e-7 | 64 | True | 1 | 0.3 | 0.849 | 0.833 |

*Learning rate was sampled from a logarithmic uniform distribution. All other variables are sampled from a categorical distribution.

[†]Integer values correspond to the first "unit" of the backbone architecture which was unfrozen for training, with 0 meaning the entire backbone is trainable. We define the choices 1, 2, 3, and 4 for each of the backbones as follows: ResNet50 Layers 1, 2, 3, and 4; DenseNet121 Dense Blocks 1, 2, 3, and 4; ViTB32 Encoder Layers 3, 6, 9, and 12.

[‡]Micro-averaged AUC over all density categories.

**Training configuration details:** Training was done with the Adam optimizer (5) and the learning rate degraded by half when no improvement in training loss is seen for 10 epochs. Training is stopped when no improvement in validation loss is observed for 25 epochs. During training, images were augmented with random cropping, rotation, and brightness/contrast jitter. In testing, images were center-cropped.

**Jud method implementation details:** Jud et. al collect experimental breast ultrasound and mammography images for 93 prospectively-recruited patients. As their images were not collected clinically, there are no imaging artifacts present. Thus, our *curated* validation set most closely reflects the development and testing environment of Jud et. al. 16-bin gray-level image histograms were calculated according to Jud et al's definition, evenly-spaced based on grayscale pixel value: [0, 15], [16, 31], etc. (3). We compared performance with and without per-image min-max normalization of gray-levels prior to binning. Models were optimized over five-fold cross-validation (by woman) on the curated validation set via grid search (see **Supplemental Table 2**). Predictions were arithmetically averaged for each woman to generate a final breast density prediction. Final linear (logistic regression) and non-linear (MLP or random forest) models were chosen based on performance on the curated validation set

**Statistical analysis step 1:** Micro-averaging of the AUROC measures how well the AI model predicts breast density when performance is weighted equivalently for all ground-truth breast density categories. Micro-averaging weights all instances equally in metric computation, creating a global metric natively. Macro-averaging calculates class-wise metrics, then takes a global average. Subgroup performance by micro-averaged AUROC is established solely on the testing set. 95% confidence intervals for AUROC are estimated using DeLong's method (6, 7). All subgroups are defined by metadata collected at time-of-exam or by cancer status collected from the Hawai'i Tumor Registry.

**Statistical analysis step 2:** Logistic regression models were constructed using 3-fold cross-validation, split by woman, on the case-control matched testing set. Logistic regression models were constructed and compared for two different measures of mammographic breast density: clinical BI-RADS breast density and BI-RADS breast density predicted by the best-performing AI model (DenseNet121). For each woman, 100 data points were simulated by independent random sampling from the predicted breast density distribution. A logistic regression was fit on all 107,400 simulated points. For odds ratio estimation, a single density was randomly sampled from the predicted distribution for each woman.

*Table 2.* Hyperparameter search space for all machine learning (ML) methods from the 16 gray-level bins defined in (3). Hyperparameters were chosen based on 5-fold cross validation on the clean validation set.

| ML Models | Chosen Value | | Clean Validation AUC[‡] |
|---|---|---|---|
| | **Selected Hyperparameters** | **Normalization** | |
| Logistic Regression | C: 10<br>Regularization method: L1<br>Solver: Saga | Yes | 0.638 |
| MLP | Activation: ReLU<br>Learning rate: 0.0001<br>Hidden layer size: 512<br>Optimizer: Adam<br>LR Schedule: Constant | Yes | 0.741 |
| Random Forest | Max tree depth: None<br>Min samples per leaf: 1<br>Min samples per split: 2<br>Number of trees: 200 | No | 0.965 |

[‡]Micro-averaged AUC over all density categories.

**Inclusion/exclusion criteria-specific counts.** 108 and 1,441 women were excluded from the testing and development sets, respectively, due to all US images being captured as part of a biopsy procedure. Of the 14,083 women remaining, 17 women were excluded due to exclusion of all their BUS images being invalid. 459,161 images were retrieved from all included women. After exclusion of images captured as part of a biopsy procedure (n = 51, 049) and invalid images (n = 4,119) and splitting dual-view images (n = 1,715), 407,423 images remained. After removal of elastography and Color Doppler scans from the testing set, 405,120 total BUS images remained.

*Table 3.* Odds ratios for cancer risk logistic regression models presented in the main text. Geras Density refers to density values obtained from the classifier presented in (1).

|  | Odds Ratio (95% CI) | | |
| --- | --- | --- | --- |
|  | **Clinical Density** | **BUS Density** | **Geras Density** |
| Fatty/A | 1.24 (0.87, 1.77) | 1.07 (0.74, 1.55) | 1.15 (0.80, 1.64) |
| Scattered/B | 1.00 (Reference) | 1.00 (Reference) | 1.00 (Reference) |
| Heterogeneous/C | 1.11 (0.63, 1.97) | 1.11 (0.66, 1.86) | 1.19 (0.70, 2.04) |
| Dense/D | 1.54 (1.31, 1.81) | 1.50 (1.28, 1.76) | 1.51 (1.29, 1.77) |
| Age* | 0.56 (0.19, 1.65) | 1.12 (0.31, 4.09) | 0.96 (0.40, 2.30) |

*Age is standardized to unit variance and zero mean.

*Table 4*. Subgroup performance statistics for the DenseNet121, logistic regression, and random forest models.

| | | DenseNet121 | | Logistic Reg | | Random Forest | |
|---|---|---|---|---|---|---|---|
| | | **Per-Image** | **Per-Patient** | **Per-Image** | **Per-Patient** | **Per-Image** | **Per-Patient** |
| | **AUC*** | 0.840 | 0.854 | 0.771 | 0.795 | 0.759 | 0.794 |
| | **(95% C.I.)** | (0.838, 0.842) | (0.842, 0.866) | (0.768, 0.774) | (0.781, 0.809) | (0.756, 0.762) | (0.780, 0.808) |
| by US machine manufacturer | PHILIPS | 0.838 (0.836, 0.841) | 0.853 (0.839, 0.866) | 0.766 (0.763, 0.770) | 0.789 (0.773, 0.806) | 0.761 (0.758, 0.764) | 0.798 (0.782, 0.814) |
| | ATL | 0.870 (0.863, 0.876) | 0.864 (0.829, 0.900) | 0.809 (0.801, 0.817) | 0.831 (0.789, 0.872) | 0.749 (0.740, 0.758) | 0.781 (0.737, 0.826) |
| | TOSHIBA‡ | 0.883 (0.862, 0.905) | 0.912 (0.870, 0.954) | 0.814 (0.788, 0.841) | 0.856 (0.809, 0.903) | 0.778 (0.750, 0.805) | 0.822 (0.771, 0.873) |
| | SIEMENS‡ | 0.718 (0.692, 0.744) | 0.793 (0.731, 0.855) | 0.699 (0.672, 0.727) | 0.739 (0.668, 0.811) | 0.687 (0.659, 0.715) | 0.736 (0.668, 0.804) |
| by binned age (years) at US exam | < 50 | 0.855 (0.852, 0.858) | 0.875 (0.859, 0.891) | 0.754 (0.749, 0.758) | 0.792 (0.772, 0.814) | 0.744 (0.740, 0.749) | 0.794 (0.773, 0.815) |
| | [50, 65) | 0.825 (0.821, 0.829) | 0.843 (0.823, 0.863) | 0.775 (0.770, 0.779) | 0.795 (0.772, 0.818) | 0.765 (0.761, 0.770) | 0.793 (0.771, 0.815) |
| | ≥ 65 | 0.826 (0.820, 0.833) | 0.818 (0.786, 0.851) | 0.809 (0.802, 0.816) | 0.793 (0.757, 0.829) | 0.787 (0.780, 0.794) | 0.793 (0.758, 0.828) |
| by cancer status | Benign | 0.844 (0.841, 0.846) | 0.854 (0.841, 0.867) | 0.769 (0.767, 0.772) | 0.793 (0.778, 0.809) | 0.759 (0.756, 0.762) | 0.793 (0.777, 0.808) |
| | Malignant | 0.822 (0.816, 0.828) | 0.853 (0.825, 0.882) | 0.780 (0.773, 0.787) | 0.803 (0.769, 0.837) | 0.758 (0.751, 0.766) | 0.801 (0.767, 0.834) |
| by US exam BI-RADS category | BI-RADS 1 | 0.826 (0.819, 0.833) | 0.858 (0.829, 0.887) | 0.785 (0.777, 0.792) | 0.812 (0.779, 0.845) | 0.767 (0.760, 0.775) | 0.810 (0.778, 0.843) |
| | BI-RADS 2 | 0.831 (0.828, 0.834) | 0.841 (0.825, 0.858) | 0.752 (0.748, 0.756) | 0.777 (0.758, 0.797) | 0.740 (0.736, 0.744) | 0.775 (0.756, 0.795) |
| | BI-RADS 3 | 0.868 (0.864, 0.873) | 0.879 (0.859, 0.900) | 0.808 (0.803, 0.814) | 0.820 (0.794, 0.846) | 0.799 (0.793, 0.804) | 0.824 (0.798, 0.849) |

*AUC is a micro-averaged one vs. rest AUROC over all density categories, except for density subgroups. For density subgroups, one vs. rest AUC for each category is shown. 95% confidence intervals are calculated using DeLong's method (1, 2).

‡Denotes subgroups for which there are fewer than 100 unique women.

| | | Predicted Classification ||
| :---: | :---: | :---: | :---: |
| | | **A/B** | **C/D** |
| **True Classification** | **A/B** | 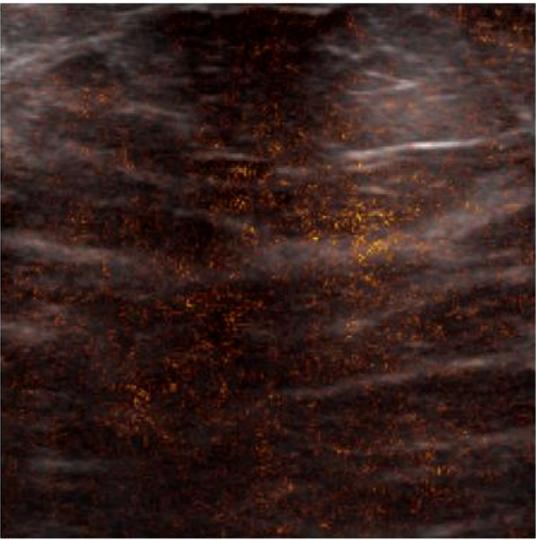 | 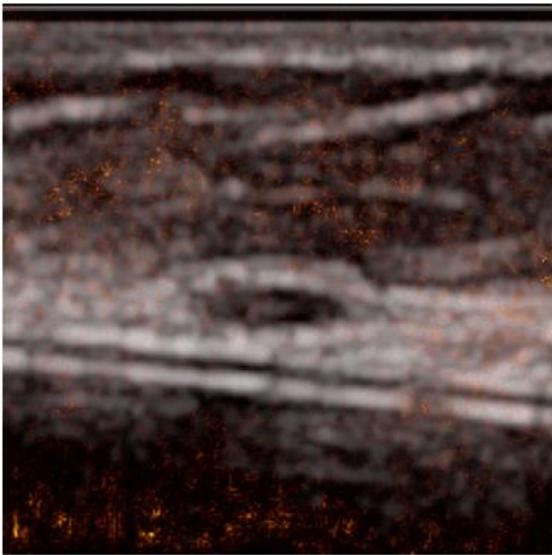 |
| | | P(B) = 0.560<br>True density A | P(D) = 0.727<br>True density B |
| | **C/D** | 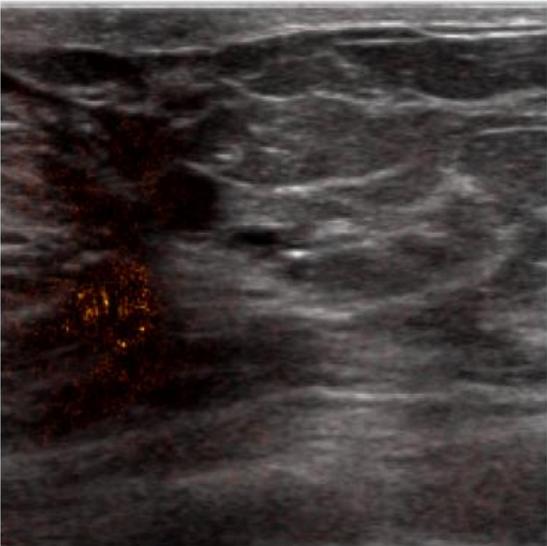 | 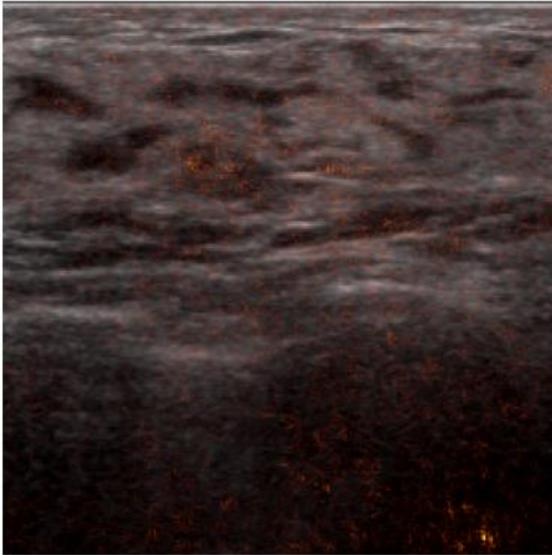 |
| | | P(B) = 0.482<br>True density D | P(D) = 0.661<br>True density D |

*Figure 1* Confusion matrix style visualization of integrated gradient maps for model predictions. All images are from the unseen test set.